\documentclass[12pt]{article} \usepackage{amssymb,latexsym}
\usepackage{amsmath}
 
\columnwidth\textwidth

\begin{document}

\newtheorem{df}{Definition} \newtheorem{thm}{Theorem} \newtheorem{lem}{Lemma}
\newtheorem{rl}{Rule} \newtheorem{assump}{Assumption}
\begin{titlepage}
 
\noindent
 
\begin{center} {\LARGE The phenomenon of state reduction} \vspace{1cm}

P. H\'{a}j\'{\i}\v{c}ek\\ Institute for Theoretical Physics \\ University of
Bern \\ Sidlerstrasse 5, CH-3012 Bern, Switzerland \\ hajicek@itp.unibe.ch

\vspace{1cm}

June 2016 \\

\vspace{1cm}
 
 
\vspace*{2cm}
 
\nopagebreak[4]
 
\begin{abstract} A theory of quantum measurement was introduced some time ago
  that was based on the notion of the so-called separation status. This
  separation status had a spatial, local character so that the theory worked
  only in special cases. Nevertheless, it enabled a description of state
  reduction process that was specific in where, when and under which objective
  conditions the process occurs and that preserved the unitary transformation
  symmetry of quantum mechanics. Now, in the accompanying paper
  (arxiv:1411.5524), a completely general mathematical definition of the
  status is given and analysed. The present paper reformulates the theory of
  state reduction accordingly. A general mathematical form of the process is
  postulated and illustrated by examples of Stern-Gerlach experiment and of
  screening.
\end{abstract}

\end{center}

\end{titlepage}

\section{Introduction} In papers \cite{hajicekC2} and \cite{hajicekC3}, we
have constructed an interpretation of quantum mechanics by defining objective
properties of quantum systems as those that are uniquely determined by
preparations and by viewing classical properties as certain special properties
of high-entropy quantum states of many-particle systems. The interpretation
has been called ``Realism-Completeness-Universality'' (RCU) Interpretation.

Accordingly, quantum states are objective properties of individual quantum
systems. Hence, as in all interpretations that associate states with
individual quantum systems, a well-known additional difficulty for the quantum
theory of measurement emerges (see, e.g., \cite{peres}, p.\ 374 and
\cite{ballent}, Section 9.2): the application of Schr\"{o}dinger equation to a
measurement process can result in linear superpositions of states that
correspond to different registration values. If the end state of the
measurement were associated with an individual system, then the state would
contradict the observed outcome, which is always just one of the possible
registration values (this is the so-called ``objectification'', see
\cite{BLM}). The transition of the linear-superposition state to the proper
mixture of definite-outcome states is called ``state reduction''.

There are many approaches to the problem in the literature. Some attempts
start from the assumption that the transition is not observable because the
registration of observables that would reveal the difference is either very
difficult or that such observables do not exist. One can then deny that the
transition really takes place and so assume that the objectification is only
apparent (no-collapse scenario). There are three most important no-collapse
approaches:
\begin{enumerate}
\item Quantum decoherence theory \cite{Zeh,Zurek,schloss}. The idea is that
system $S +M$ composed of a quantum system $S$ and an apparatus $M$ cannot be
isolated from environment $E$. Then the unitary evolution of $S + M + E$ leads
to a non-unitary evolution of $S + M$ that can erase all correlations and
interferences from $S + M$ hindering the objectification
\cite{Zeh,Zurek,schloss} (see discussion in Refs.\
\cite{d'Espagnat,Ghirardi,bub,BLM}).
\item Superselection sectors approach \cite{Hepp,Primas,wanb}. Here, classical
properties are described by superselection observables of $M$ which commute
with each other and with all other observables of $M$. Then, the state of $M$
after the measurement is equivalent to a suitable proper mixture.
\item Modal interpretation \cite{bub}. One assumes that there is a subset of
orthogonal-projection observables that, first, can have determinate values in
the state of $S + M$ before the registration in the sense that the assumption
does not violate contextuality (see e.g.\ \cite{peres}, Chapter 7) and second,
that one can reproduce all important results of ordinary quantum mechanics
with the help of these limited set of observables. Thus, one must require that
the other observables are not registered. An analogous requirement can be
identified in any of the no-collapse approaches.
\end{enumerate}

Other attempts (collapse scenario) do assume that the reduction is a real
process and postulate a new dynamics that leads directly to something
analogous to the reducing transition accepting the consequence that some
measurement could disprove this postulate. An example of the collapse scenario
is known as Dynamical Reduction Program \cite{GRW,pearle}. It postulates new
universal, unique quantum dynamics that is non-linear and stochastic. Both the
unitary evolution and the state reduction result as some approximations. The
physical idea is that of spontaneous localisation: linear superpositions of
different positions spontaneously decay, either by jumps \cite{GRW} or by
continuous transitions \cite{pearle}. The form of this decay is chosen
judiciously to take a very long time for microsystems, so that the standard
quantum mechanics is a good approximation, and a very short time for
macrosystems, leading to practically immediate state reductions. In this way,
a simple explanation of the definite positions of macroscopic systems and of
the pointers of registration apparatuses is achieved.

One of the important ideas of the Dynamical Reduction Program is to make the
state reduction well-defined by choosing a particular frame for it: the
$Q$-representa\-tion. This leads to breaking of the symmetry with respect to
all unitary transformations that was not only a beautiful but also a practical
feature of standard quantum mechanics.

Another example of collapse scenario is our approach (see Refs.\
\cite{hajicek2,survey,hajicekW1,hajicek5}). Its aim is to postulate the
existence of state reductions so that it does not break the unitary symmetry,
even if it itself is a non-unitary transformation, and to formulate hypotheses
about the conditions, origin and form of state reduction.

For this approach, the notion of the so-called {\em separation status} is
instrumental. In above papers, it was defined just for spatial separations
which made the theory of state reduction valid only in some special cases. In
the accompanying paper \cite{incomplete}, a completely general definition is
now given and the present paper reformulates the theory of state reduction
accordingly. It uses a number of results from \cite{incomplete} without
introducing them anew. Hence, the present paper can only be understood if
\cite{incomplete} is at hand.

\section{Reformulation of the standard theory} This section explains the
reformulation with the help of models. It also introduces the necessary
technical tools.

\subsection{Stern-Gerlach story retold} Here, we modify the textbook
description (e.g., \cite{peres}, pp.\ 14 and 375 or \cite{ballent}, p.\ 230)
of the Stern-Gerlach experiment. There are two changes. First, we take more
seriously the role of real detectors in the experiment. The detector is
assumed to be an object with both classical and quantum model that gives
information on the registered quantum object via its classical
properties. Hence, it has to satisfy the assumptions of Ref.\ \cite{hajicekC3}
on classical properties. Second, the description is made compatible with the
consequences of the exchange symmetry for the measurement process that were
explained in Ref.\ \cite{incomplete} so that it can make use of changes of
separation status.

The original experiment measures the spin of silver atoms. A silver atom
consists of 47 protons and 61 neutrons in the nucleus and of 47 electrons
around it. This leads to some complications that can be dealt with technically
but that would obscure the ideas we are going to illustrate. To simplify, we
replace the silver atom by a neutral spin 1/2 particle.

Let the particle be denoted by $S$ and its Hilbert space by ${\mathbf H}$. Let
$\vec{\mathsf x}$ be its position, $\vec{\mathsf p}$ its momentum and
${\mathsf S}_z$ the $z$-component of its spin with eigenvectors $|j\rangle$
and eigenvalues $j \hbar/2$, where $j = \pm 1$ (see e.g.\ \cite{ballent},
Section 7.4).

Let ${\mathcal M}$ be a Stern-Gerlach apparatus with an inhomogeneous magnetic
field oriented so that it separates different $z$-components of spin of $S$
arriving there. To calculate the evolution of $S$ in the magnetic field, we
use the modified Schr\"{o}dinger equation that describes the interaction
between the particle and external field, as it is done, e.g., in \cite{peres},
p.\ 375.

Let the detector of the apparatus be a photo-emulsion film ${\mathcal D}$ with
energy threshold $E_0$. Its emulsion grains are not macroscopic in the sense
that each would contain about $10^{23}$ molecules. They contain only about
$10^{10}$ in average. Still, the chemical and thermodynamic process in them
can be described with a sufficient precision by classical chemistry and
phenomenological thermodynamics. They have classical states and classical
properties. The emulsion grains that are hit by $S$ run through a process of
change and of modification and the modification can be made directly
visible. ${\mathcal D}$ is a macroscopic object formed by such grains. Let its
classical model be $D_c$ and its quantum one be $D_q$ with Hilbert space
${\mathbf H}^{\mathcal D}$. According to our theory of classical properties in
Ref.\ \cite{hajicekC3}, the quantum states of the grains, and so of the whole
$D_q$, must be some high-entropy states. The usual description of meters by
wave functions is thus not completely adequate.

First, let $S$ be prepared at time $t_1$ in a definite spin-component state,
\begin{equation}\label{sin} |\text{in},j\rangle = |\vec{p},\Delta
\vec{p}\rangle \otimes |j \rangle\ ,
\end{equation} where $|\vec{p},\Delta \vec{p}\rangle$ is a Gaussian wave
packet with the expectation value $\vec{p}$ and variance $\Delta \vec{p}$ of
momentum. To make the mathematics easier, we shall also work with the
formalism of wave functions and kernels explained in Ref.\
\cite{incomplete}. Thus, the wave function of state (\ref{sin}) in an
arbitrary representation will be denoted by $\psi_j(\lambda)$. Let system
$D_q$ be prepared in metastable state ${\mathsf T}^{\mathcal D}$ at $t_1$. We
assume that $D_q$ consists of $N$ particles of which $N_1$ ($N_1$ can also be
zero) are indistinguishable from $S$. Hence, the kernel of ${\mathsf
T}^{\mathcal D}$ is
$$
T^{\mathcal
D}(\lambda^{(1)},\ldots,\lambda^{(N_1)},\lambda^{(N_1+1)},\ldots,\lambda^{(N)};
\lambda^{(1)\prime},\ldots,\lambda^{(N_1)\prime},\lambda^{(N_1+1)\prime},\ldots,\lambda^{(N)\prime})\
,
$$
where the function $T^{\mathcal D}$ is antisymmetric both in variables
$\lambda^{(1)},\ldots,\lambda^{(N_1)}$ and $\lambda^{(1)\prime},\ldots$,
$\lambda^{(N_1)\prime}$. The initial state of the composite $S + D_q$ then is
\begin{equation}\label{initSD1} \bar{\mathsf T}_j =
N^2_{\text{exch}}\bar{\mathsf
\Pi}^{N_1+1}_-\Bigl(\psi_j(\lambda^{(0)})\psi^*_j(\lambda^{(0)\prime})T^{\mathcal
D}(\lambda^{(1)},\ldots,\lambda^{(N)};
\lambda^{(1)\prime},\ldots,\lambda^{(N)\prime})\Bigr)\bar{\mathsf
\Pi}^{N_1+1}_-\ ,
\end{equation} where $\bar{\mathsf \Pi}^{N_1+1}_-$ denotes the
antisymmetrisation in the variables $\lambda^{(0)},\ldots,\lambda^{(N_1)}$ (or
$\lambda^{(0)\prime},\ldots,\lambda^{(N_1)\prime}$). It is an orthogonal
projection acting on Hilbert space ${\mathbf H} \otimes {\mathbf H}^{\mathcal
D}$ (see \cite{incomplete}).

We also assume that the direction of $\vec{p}$ is suitably restricted and its
magnitude respects the energy threshold $E_0$. Such states lie in the domain
of the apparatus ${\mathcal M}$, see \cite{incomplete}. According to our
theory of meters in Section 3 of \cite{incomplete}, states in the domain of
${\mathcal M}$ have a separation status before their registration by
${\mathcal M}$. Hence, state (\ref{sin}) has a separation status at $t_1$ and
so the system $S$ represents initially an individual quantum object with an
objective state. From Definition 3 in \cite{incomplete} of separation status,
it follows that
\begin{equation}\label{DSss1} \int d\lambda^{(k)}\,
\psi^*_j(\lambda^{(k)})T^{\mathcal D}(\lambda^{(1)},\ldots,\lambda^{(N_1)};
\lambda^{(1)\prime},\ldots,\lambda^{(N_1)\prime}) = 0
\end{equation} for any $k = 1,\ldots,N_1$, and
\begin{equation}\label{DSss2} \int d\lambda^{(l)\prime}\,
\psi_j(\lambda^{(l)\prime})T^{\mathcal
D}(\lambda^{(1)},\ldots,\lambda^{(N_1)};
\lambda^{(1)\prime},\ldots,\lambda^{(N1)\prime}) = 0
\end{equation} for any $l = 1,\ldots,N_1$.

To take the exchange symmetry into account, we need the following Lemma:
\begin{lem}\label{lemNPia} Let $F_n(\lambda^{(1)},\ldots,\lambda^{(N)})$, $n =
1,\ldots,K$, be $K$ functions of $N$ variables that satisfy:
\begin{enumerate}
\item Function $F_n$ is antisymmetric in the variables
$\lambda^{(1)},\ldots,\lambda^{(N_1)}$ for all $n$ and for some $N_1 < N$.
\item For some functions $\psi_j(\lambda)$, $j=1,\ldots,L$, such that $\int
d\lambda\, \psi^*_j(\lambda)\psi_j(\lambda) = 1$,
\begin{equation}\label{lem1} \int d\lambda^{(k)}\,
\psi_j(\lambda^{(k)})F_n(\lambda^{(1)},\ldots,\lambda^{(N)}) = 0
\end{equation} for all $j$, $n$ and $k = 1,\ldots,N_1$.
\item $\{F_n\}$ is an orthonormal set,
\begin{equation}\label{lem2} \int d^N\lambda\,
F^*_{n'}(\lambda^{(1)},\ldots,\lambda^{(N)})
F_n(\lambda^{(1)},\ldots,\lambda^{(N)}) = \delta_{nn'}
\end{equation} for all $n,n'$.
\end{enumerate} Let function $\bar{F}_{jn}$ of $N+1$ variables
$\lambda^{(0)},\lambda^{(1)},\ldots,\lambda^{(N)}$ be defined by
\begin{equation}\label{lem3}
\bar{F}_{jn}(\lambda^{(0)}\lambda_1,\ldots,\lambda^{(N)}) =
\frac{1}{\sqrt{N_1+1}} \sum_{k=0}^{N_1} (-1)^{kN_1} \psi_j(\lambda^{(k)})
F_n[\lambda^{(0)}\mapsto \lambda^{(k)}]\ ,
\end{equation} where
$$
F_n[\lambda^{(0)}\mapsto \lambda^{(k)}] =
F_n(\lambda^{(k+1)},\ldots,\lambda^{(N_1)},\lambda^{(0)},\ldots,\lambda^{(k-1)},\lambda^{(N_1+1)},\ldots,\lambda^{(N)})\
.
$$
Then functions $\bar{F}_{jn}$ are antisymmetric in variables
$\lambda^{(0)},\lambda^{(1)},\ldots,\lambda^{(N_1)}$ and satisfy:
\begin{equation}\label{lem4} \int d^{N+1}\lambda\,
\bar{F}^*_{jn}(\lambda^{(0)},\lambda^{(1)},\ldots,\lambda^{(N)})
\bar{F}_{jn'}(\lambda^{(0)},\lambda^{(1)},\ldots,\lambda^{(N)}) = \delta_{nn'}
\end{equation} for all $j$, $n$ and $n'$.
\end{lem} The set $\lambda^{(a)},\ldots,\lambda^{(b)}$ for any integers $a$
and $b$ is empty if $a > b$ and contains all entries $\lambda^{(c)}$ for $a
\leq c \leq b$ in the increasing index order if $a \leq b$.\par\noindent {\bf
Proof} Function $\bar{F}_{jn}$ is antisymmetric because $F_n$ is and the sum
in (\ref{lem3}) contains already exchanges of $\lambda^{(0)}$ and
$\lambda^{(k)}$ for all $k > 0$ with the proper signs (see Eq. (12) of
\cite{incomplete}). To show Eq.\ (\ref{lem4}), we substitute Eq.\ (\ref{lem3})
into the right-hand side of Eq.\ (\ref{lem4}):
\begin{multline*} \int d^{N+1}\lambda\, \bar{F}^*_{jn'}\bar{F}_{jn} =
\frac{1}{N_1+1} \int d^{N+1}\lambda \sum_{k=0}^{N_1}
(-1)^{kN_1}\sum_{l=0}^{N_1} (-1)^{lN_1} \\ \times \psi_j(\lambda^{(k)})
F_n[\lambda^{(0)}\mapsto \lambda^{(k)}]
\psi^*_j(\lambda_l)F^*_{n'}[\lambda^{(0)}\mapsto \lambda^{(l)}]\ .
\end{multline*} The terms
$$
\int d^{N+1}\lambda\, \psi_j(\lambda^{(k)}) F_n[\lambda^{(0)}\mapsto
\lambda^{(k)}] \psi^*_j(\lambda^{(1l)})F^*_{n'}[\lambda^{(0)}\mapsto
\lambda^{(l)}]
$$
vanish for any $k \neq l$ because of Eq.\ (\ref{lem1}). The remaining terms
$$
\int d^{N+1}\lambda\, \psi_j(\lambda^{(k)}) F_n[\lambda^{(0)}\mapsto
\lambda^{(k)}] \psi^*_j(\lambda^{(k)})F^*_{n'}[\lambda^{(0)}\mapsto
\lambda^{(k)}]
$$
are equal to $\delta_{nn'}$ for all $k$ because of the normalisation of
$\psi_j$ and Eq.\ ({\ref{lem2}), {\bf QED}.

State (\ref{initSD1}) has then the following kernel:
\begin{multline}\label{kernbarTj}
\bar{T}_j(\lambda^{(0)},\ldots,\lambda^{(N)};
\lambda^{(0)\prime},\ldots,\lambda^{(N)\prime}) =
\frac{1}{N_1+1}\sum_{k=0}^{N_1} (-1)^{kN_1}\sum_{l=0}^{N_1} (-1)^{lN_1}
\psi_j(\lambda^{(k)})\psi^*_j(\lambda^{(l)\prime}) \\ T^{\mathcal
D}(\lambda^{(k+1)},\ldots,\lambda^{(N_1)},\lambda^{(0)},\ldots,\lambda^{(k-1)},\lambda^{(N_1+1)},\ldots,\lambda^{(N)};
\\
\lambda^{(l+1)\prime},\ldots,\lambda^{(N_1)\prime},\lambda^{(0)\prime},\ldots,\lambda^{(l-1)\prime},\lambda^{(N_1+1)\prime},\ldots,\lambda^{(N)\prime})\
.
\end{multline} Kernel $\bar{T}_j$ can be shown to be antisymmetric in
variables $\lambda^{(0)},\ldots,\lambda^{(N_1)}$ and
$\lambda^{(0)\prime},\ldots,$ $\lambda^{(N_1)\prime}$ and to have trace equal
1 by the same methods as those used to prove Lemma \ref{lemNPia}. Eqs.\
(\ref{DSss1}) and (\ref{DSss2}) expressing the separation status of
$|\psi\rangle$ play an important role in the derivation of formula
(\ref{kernbarTj}).

The initial state of $S + D_q$ does not contain any modified emulsion
grains. Such states, if extremal, form a subspace of the Hilbert space
$\bar{\mathsf \Pi}^{N_1+1}_-({\mathbf H} \otimes {\mathbf H}^{\mathcal D})$ of
$S + D_q$. Let us denote the projection to this subspace by $\bar{\mathsf
\Pi}[\emptyset]$. Thus, we have
\begin{equation}\label{nograins} tr(\bar{\mathsf T}_j\bar{\mathsf
\Pi}[\emptyset]) = 1\ .
\end{equation}

The process of registration includes the interaction of $S$ with the magnetic
field and with system $D_q$ as well as the resulting modification of the
emulsion grains. We assume that meter ${\mathcal M}$ is {\em ideal}: each copy
of $S$ that arrives at the emulsion $D_q$ modifies at least one emulsion
grain.

The registration is assumed to be a quantum evolution described by a unitary
group $\bar{\mathsf U}(t)$, the so-called measurement coupling. We assume that
$\bar{\mathsf U}(t)$ commutes with $\bar{\mathsf \Pi}^{N_1+1}_-$, see Section
5 of \cite{incomplete}. Let $t_2$ be the time at which the modification of the
hit grains is finished and let $\bar{\mathsf U} = \bar{\mathsf U}(t_2 -
t_1)$. We are going to derive some important properties of $\bar{\mathsf
U}{\mathsf T}_j\bar{\mathsf U}^\dagger$, and for this we need a technical
trick that transforms calculations with kernels into that with wave functions.

Let
\begin{equation}\label{specdecTD} {\mathsf T}^{\mathcal D} = \sum_n a_n
|n\rangle \langle n|
\end{equation} be the spectral decomposition of ${\mathsf T}^{\mathcal
D}$. Then, $0 \leq a_n \leq 1$ for each $n \in {\mathbb N}$ and $\sum_n a_n =
1$. In $\lambda$-representation, state $|n\rangle$ has the wave function
$\varphi_n(\lambda^{(1)},\ldots,\lambda^{(N)})$.

Eqs.\ (\ref{kernbarTj}) and (\ref{specdecTD}) imply that
\begin{equation}\label{sdbarTj} \bar{T}_j(\lambda^{(0)},\ldots,\lambda^{(N)};
\lambda^{(0)\prime},\ldots,\lambda^{(N)\prime}) = \sum_na_n
\bar{\Psi}_{jn}(\lambda^{(0)},\ldots,\lambda^{(N)})\bar{\Psi}^*_{jn}(\lambda^{(0)\prime},\ldots,\lambda^{(N)\prime})\
,
\end{equation} where
\begin{equation}\label{compinvec}
\bar{\Psi}_{jn}(\lambda^{(0)},\ldots,\lambda^{(N)}) = \frac{1}{\sqrt{N_1+1}}
\sum_{k=0}^{N_1} (-1)^{kN_1}
\psi_j(\lambda^{(k)})\varphi_n[\lambda^{(0)}\mapsto \lambda^{(k)}]\ .
\end{equation}
\begin{lem}\label{lemsdbarTj} Eq.\ (\ref{sdbarTj}) is the spectral
decomposition of state $\bar{\mathsf T}_j$.
\end{lem} {\bf Proof} Conditions (\ref{DSss1}) and (\ref{DSss2}) on
$|\psi_j\rangle$ and $\bar{\mathsf T}^{\mathcal D}$ imply
$$
\sum_n a_n \int d\lambda^{(k)} \int d\lambda^{(l)\prime}\,
\psi^*_j(\lambda^{(k)})\psi_j(\lambda^{(k)\prime})\,
\varphi_n(\lambda^{(1)},\ldots,\lambda^{(N)})
\varphi_n^*(\lambda^{(1)\prime},\ldots,\lambda^{(N)\prime}) = 0$$ for all $k=
1,\dots,N_1$. However, the integral defines a positive kernel
$$
K_n(\lambda^{(1)},\ldots,\lambda^{(k-1)}\lambda^{(k+1)},\ldots,\lambda^{(N)};\lambda^{(1)\prime},\ldots,\lambda^{(k-1)\prime}\lambda^{(k+1)\prime},\ldots,\lambda^{(N)\prime})
$$
for each $n$ and a sum with positive coefficients of such kernels can be zero
only if each such kernel itself vanishes. Hence, we have
\begin{equation}\label{wfSDss} \int d\lambda_k\,
\psi^*(\lambda^{(k)})\varphi_n(\lambda^{(1)},\ldots,\lambda^{(N)}) = 0
\end{equation} for each $n$ and all $k = 1,\ldots,N$.

From Lemma \ref{lemNPia}, it then follows now that
$$
\langle \bar{\Psi}_{jn}|\bar{\Psi}_{jn'}\rangle = \delta_{nn'}\ .
$$
This implies Lemma \ref{lemsdbarTj}, {\bf QED}.

A simple consequence of Lemma \ref{lemsdbarTj} is the following. Combining
Eqs.\ (\ref{nograins}) and (\ref{sdbarTj}), we obtain
$$
tr\Bigl(\sum_na_n |\bar{\Psi}_{jn}\rangle \langle\bar{\Psi}_{jn}|\bar{\mathsf
\Pi}[\emptyset] \Bigl) = \sum_na_n tr\Bigl((\bar{\mathsf
\Pi}[\emptyset]|\bar{\Psi}_{jn}\rangle) (\langle\bar{\Psi}_{jn}|\bar{\mathsf
\Pi}[\emptyset]) \Bigl) = 1\ .
$$
But operator $\bar{\mathsf \Pi}[\emptyset]|\bar{\Psi}_{jn}\rangle
\langle\bar{\Psi}_{jn}|\bar{\mathsf \Pi}[\emptyset]$ is positive so that its
trace must be non-negative. As the sum of $a_n$'s is already 1, we must have
$$
tr(\bar{\mathsf \Pi}[\emptyset]|\bar{\Psi}_{jn}\rangle
\langle\bar{\Psi}_{jn}|\bar{\mathsf \Pi}[\emptyset]) = 1
$$
or
$$
\langle\bar{\mathsf
\Pi}[\emptyset]|\bar{\Psi}_{jn}|\bar{\Psi}_{jn}|\bar{\mathsf
\Pi}[\emptyset]\rangle = 1
$$
for each $n$. However,
$$
|\bar{\Psi}_{jn}\rangle = \bar{\mathsf \Pi}[\emptyset]|\bar{\Psi}_{jn}\rangle
+ ({\mathsf 1} - \bar{\mathsf \Pi}[\emptyset])|\bar{\Psi}_{jn}\rangle
$$
and
$$
\langle \bar{\mathsf \Pi}[\emptyset]|\bar{\Psi}_{jn}|({\mathsf 1} -
\bar{\mathsf \Pi}[\emptyset])|\bar{\Psi}_{jn}\rangle = 0
$$
so that
$$
1 = \langle \bar{\Psi}_{jn}|\bar{\Psi}_{jn}\rangle = \langle \bar{\mathsf
\Pi}[\emptyset]|\bar{\Psi}_{jn}|\bar{\mathsf
\Pi}[\emptyset]|\bar{\Psi}_{jn}\rangle + \langle ({\mathsf 1} - \bar{\mathsf
\Pi}[\emptyset])|\bar{\Psi}_{jn}|({\mathsf 1} - \bar{\mathsf
\Pi}[\emptyset])|\bar{\Psi}_{jn}\rangle\ .
$$
Hence,
\begin{equation}\label{PiPsi} \bar{\mathsf
\Pi}[\emptyset]|\bar{\Psi}_{jn}\rangle = |\bar{\Psi}_{jn}\rangle\ .
\end{equation}

Let us now return to the time evolution of $\bar{\mathsf T}_j$ within
$\bar{\mathsf \Pi}^{N_1+1}_-({\mathbf H} \otimes {\mathbf H}^{\mathcal D})$
from $t_1$ to $t_2$. System $S + D_q$ is composed of two subsystems, $S'$ and
$D_q'$, $S'$ containing $S$ and all $N_1$ particles of $D_q$ that are
indistinguishable from $S$. Then, $\bar{\mathsf \Pi}^{N_1+1}_-({\mathbf H}
\otimes {\mathbf H}^{\mathcal D}) = ({\mathbf H})^{N_1+1}_- \otimes {\mathbf
H}^{\mathcal D\prime}$. The evolution defines states $\bar{\mathsf
T}_{j}(t_2)$ of $S + D_q$ by:
\begin{equation}\label{evolTt1} \bar{U}\bar{\mathsf T}_j\bar{U}^\dagger =
\bar{\mathsf T}_{j}(t_2)\ .
\end{equation} Evolution $\bar{\mathsf U}$ includes a thermodynamic relaxation
of $S + D_q$ and a loss of separation status of $S$ if $S$ and $S'$ do not
coincide. Thus, in general, quantum system $S$ does not represent an
individual quantum object after the registration. The individual states that
could be ascribed to $S$ as its objective properties are not well defined (see
\cite{incomplete}) at $t = t_2$. We can say that they do not exist. However,
the whole composite $S + D_q$ is a quantum object, prepared in the measurement
experiment, hence one can consider its individual states as its objective
properties (see \cite{hajicekC3}).

Accordingly, states $\bar{\mathsf T}_{j}(t_2)$ also describe the modified
emulsion grains, which can be called {\em detector signals}. The signals are
concentrated within two strips of the film, each strip corresponding to one
value of $j$. The two space regions, $R_+$ and $R_-$, of the two strips are
sufficiently separated and help to determine, in the present case, what is
generally called a pointer observable: the occurrence of a modified emulsion
grain within $R_+$ or $R_-$. Let the projections onto the subspaces of
$({\mathbf H})^{N_1+1}_- \otimes {\mathbf H}^{\mathcal D\prime}$ containing
the corresponding extremal states be $\bar{\mathsf \Pi}[R_j]$.

We avoid specifying $\bar{\mathsf U}(t)$ e.g.\ by writing the Hamiltonian of
system $S + {\mathcal D}_q$. Instead, we express the condition that the meter
registers ${\mathsf S}_z$ through properties of end states ${\mathsf
T}_{j}(t_2)$ as follows:
\begin{equation}\label{registcond1} tr\Bigl(\bar{\mathsf \Pi}[R_j]\bar{\mathsf
T}_{k}(t_2)\Bigr) = \delta_{jk}\ .
\end{equation}

If we substitute Eqs.\ (\ref{evolTt1}) and (\ref{sdbarTj}) into
(\ref{registcond1}), we obtain
$$
\sum_na_n tr(\bar{U}|\bar{\Psi}_{kn}\rangle \langle
\bar{\Psi}_{kn}|\bar{U}^\dagger\bar{\mathsf \Pi}[R_j]) = \delta_{jk}\ .
$$
By the same argument as that leading to formula (\ref{PiPsi}), we then have
\begin{equation}\label{PiDjPsik} \bar{\mathsf
\Pi}[R_j]|\bar{\Psi}_{kn}(t_2)\rangle =
\delta_{jk}|\bar{\Psi}_{kn}(t_2)\rangle\ ,
\end{equation} where
$$
|\bar{\Psi}_{kn}(t_2)\rangle = \bar{U}|\bar{\Psi}_{kn}\rangle\ .
$$
Hence, the state $\bar{U}|\bar{\Psi}_{kn}\rangle$ contains modified emulsion
grains in the region $R_k$ and no such grains in the region $R_l$ for each $n$
and $l \neq k$.

Suppose next that the initial state of $S$ at $t_1$ is
\begin{equation}\label{initS2} |\text{in}\rangle = \sum_j c_j
|\text{in},j\rangle
\end{equation} with
$$
\sum_j|c_j|^2 = 1\ .
$$

The linearity of $\bar{\mathsf U}$ implies the following form of the
corresponding end state $\bar{\mathsf T}(t_2)$:
\begin{multline}\label{endstat2} \bar{\mathsf T}(t_2) =
N^2_{\text{exch}}\bar{\mathsf U}\bar{\mathsf \Pi}^{N_1+1}_-\left[
\left(\sum_jc_j|\text{in},j \rangle\right) \left(\sum_{j'}c^*_{j'}\langle
\text{in},j'|\right)\otimes {\mathsf T}^{\mathcal D}\right]\bar{\mathsf
\Pi}^{N_1+1}_-\bar{\mathsf U}^\dagger \\ = \sum_{jj'}c_jc^*_{j'} \bar{\mathsf
T}_{jj'}(t_2)\ ,
\end{multline} Operators $\bar{\mathsf T}_{jj'}(t_2)$ act on the Hilbert space
$\bar{\mathsf \Pi}^{N_1+1}_-({\mathbf H} \otimes {\mathbf H}^{\mathcal D})$ of
$S + D_q$ and are defined by
\begin{equation}\label{barTjj'} \bar{\mathsf T}_{jj'}(t_2) =
N_{\text{exch}}\bar{\mathsf U}\bar{\mathsf \Pi}^{N_1+1}_-(|\text{in},j \rangle
\langle \text{in},j'|\otimes {\mathsf T}^{\mathcal D})\bar{\mathsf
\Pi}^{N_1+1}_-\bar{\mathsf U}^\dagger\ .
\end{equation} They are state operators only for $j' = j$. Eqs.\
(\ref{evolTt1}) and (\ref{initSD1}) imply that
$$
{\mathsf T}_{jj}(t_2) = {\mathsf T}_{j}(t_2)\ .
$$

If we substitute the spectral decomposition (\ref{specdecTD}) of ${\mathsf
T}^{\mathcal D}$ into Eq.\ (\ref{barTjj'}), we obtain for the kernel of
operator ${\mathsf T}_{jj'}(t_2)$
\begin{multline*} T_{jj'}(t_2) = \sum_n a_n \bar{\mathsf U} \\ \times
\Bigl(\sum_{k=0}^{N_1}(-1)^{N_1k}\psi_j(\lambda^{(k)})\varphi_n(\lambda^{(k+1)},\ldots,\lambda^{(N_1)},\lambda^{(0)},\ldots,\lambda^{(k-1)},\lambda^{(N_1+1)},\ldots,\lambda^{(N)})\Bigr)
\\ \times \Bigl(\sum_{l=0}^{N_1}
(-1)^{N_1l}\psi^*_{j'}(\lambda^{(l)}\prime)\varphi^*_n(\lambda^{(l+1)\prime},\ldots,\lambda^{(N_1)\prime},\lambda^{(0)\prime},\ldots,\lambda^{(l-1)\prime},\lambda^{(N_1+1)\prime},\ldots,\lambda^{(N)\prime})\Bigr)\bar{\mathsf
U}^\dagger \\ = \sum_n a_n \Bigl(\bar{\mathsf
U}\bar{\Psi}_{jn}(\lambda^{(0)},\ldots,\lambda^{(N)})\Bigr)\Bigl(\bar{\Psi}^*_{j'n}(\lambda^{(0)\prime},\ldots,\lambda^{(N)\prime})\bar{\mathsf
U}^\dagger\Bigr)\ ,
\end{multline*} or
\begin{equation}\label{decbarTjj'} {\mathsf T}_{jj'}(t_2) = \sum_n a_n
|\bar{\Psi}_{jn}(t_2)\rangle \langle \bar{\Psi}_{j'n}(t_2)|\ .
\end{equation} Eq.\ (\ref{decbarTjj'}) is, of course, not the spectral
decomposition of ${\mathsf T}_{jj'}(t_2)$ because this operator is not
self-adjoint, but it can be used to show that Eq.\ (\ref{PiDjPsik}) implies:
\begin{equation}\label{compoutvec} tr\Bigl(\bar{\mathsf
\Pi}[R_k]|\bar{\Psi}_{jn}(t_2)\rangle \langle \bar{\Psi}_{j'n}(t_2)|\Bigr) =
\delta_{kj}\delta_{kj'}\ .
\end{equation} Then, because of the orthonormality of state vectors
$|\bar{\Psi}_{jn}(t_2)\rangle$, it follows that
\begin{equation}\label{registcond3} tr\Bigl(\bar{\mathsf \Pi}[R_j]\bar{\mathsf
T}_{kl}(t_2)\Bigr) = \delta_{jk}\delta_{jl}
\end{equation} and
\begin{equation}\label{PRC} tr\Bigl(\bar{\mathsf \Pi}[R_j] \bar{\mathsf
T}(t_2)\Bigr) = |c_j|^2\ .
\end{equation} The significance of Eq.\ (\ref{PRC}) is that the modified
grains will be found in the strip $j$ with the probability given by the Born
rule for registering the spin $j$ in the state (\ref{initS2}).

Eq.\ (\ref{endstat2}) can be written as
\begin{equation}\label{decom101} \bar{\mathsf T}(t_2) = \bar{\mathsf
T}_{\text{end}1} + \bar{\mathsf T}_{\text{end}0}\ ,
\end{equation} where
\begin{equation}\label{decom102} \bar{\mathsf T}_{\text{end}1} = \sum_j
|c_j|^2 \bar{\mathsf T}_j(t_2)\ ,\quad \bar{\mathsf T}_{\text{end}0} =
\sum_{j\neq j'} c_j c^*_{j'} \bar{\mathsf T}_{jj'}(t_2)\ .
\end{equation} It follows that
\begin{equation}\label{decom103} tr(\bar{\mathsf T}_{\text{end}1}) = 1\ ,\quad
tr(\bar{\mathsf T}_{\text{end}0}) = 0\ .
\end{equation} Eq.\ (\ref{decom102}) says that $\bar{\mathsf T}_{\text{end}1}$
is a convex combination of quantum states that differ from each other by
expectation values of operator $\bar{\mathsf \Pi}[R_j]$.

Finally, we have to analyse more closely what is observed in Stern-Gerlach
experiment. The basic fact is that there are modified emulsion grains at some
definite positions at the film after each registration. This is represented by
definite states of the classical model $D_c$ of the film. A basic assumption
about classical models is that their states are objective, that is, they exist
before being observed and the observation only reveals them (see
\cite{hajicekC3}). A state $T_c$ of $D_c$ can be described by specifying the
positions of the modified grains. Then we can express the fact that the
modified grains lie in strip $R_j$ by the classical state represented by
expression $T_c \subset R_j$. Quantum mechanics can only give us the
probabilities ${\mathrm P}(T_c)$ that state $T_c$ is observed:
$$
{\mathrm P}(T_c) = tr\Bigl(\bar{\mathsf T}(t_2)\bar{\mathsf \Pi}[R_j]\Bigr)\ .
$$

According to Minimum Interpretation, state $\bar{\mathsf T}(t_2)$ just
describes the statistics of the ensemble of particular measurements on system
$S + D_q$ and does not refer to anything existing before the registration and
concerning each individual system.

According to RCU Interpretations, state $\bar{\mathsf T}(t_2)$ is a property
referring directly to each individual composite system $S + D_q$ immediately
before the registration. Moreover, two quantum states $\bar{\mathsf
T}_j(t_2)$, $j = 1,2$, are in a bijective relation with two classical states
$T_c \subset R_j$, $j = 1,2$. The observation that the classical state of
$D_c$ is $T_c \subset R_j$ implies, therefore, that the quantum state of $S +
D_q$ must be $\bar{\mathsf T}_j(t_2)$ already before the (classical)
observation. Hence, the state of the individual composite system $S + D_q$
immediately before the registration must be a proper mixture of states
$\bar{\mathsf T}_j(t_2)$ each of which has a definite value of $j$:
\begin{equation}\label{endst2} \sum_j\ +_s\ |c_j|^2 \bar{\mathsf T}_{j}(t_2)
\end{equation} instead of (\ref{endstat2}) that results by unitary, linear
evolution law of quantum mechanics. Observe that the transition from state
(\ref{endstat2}) to (\ref{endst2}) is non-linear, but it preserves the norm of
the state. This additional ``evolution'' from state (\ref{endstat2}) to state
(\ref{endst2}) that must then be caused in some way by the registration, is
the state reduction.

The present subsection was rather technical because it was to describe
registrations in a way that was in agreement with the results of
\cite{incomplete} on the influence of exchange symmetry on registration and of
\cite{hajicekC3} on classical states. In particular, we avoided the need for
the definite state of the registered system after the registration as it is
usually assumed, see, e.g., \cite{BLM}.

\subsection{Screen} Screens are used in most preparation procedures. For
example, in optical experiments \cite{RSH}, polarisers, such as Glan-Thompson
ones, are employed. A polariser contains a crystal that decomposes the coming
light into two orthogonal-polarisation parts. One part disappears inside an
absorber and the other is left through. Similarly, the Stern-Gerlach
experiment can be modified so that the beam corresponding to spin down is
blocked out by an absorber and the other beam is left through. In the
interference experiment \cite{tono}, there are several screens, which are just
walls with openings. Generally, a screen is a macroscopic body that decomposes
the incoming, already prepared, beam into one part that disappears inside the
body and the other that goes through.

Here, a simple model of screen is constructed and its physics is studied. Let
the particle $S$ interacting with the screen have mass $\mu$ and spin 0 and
the screen have the following geometry:

The screen is at $x_3 = 0$ and the half-spaces $x_3 < 0$ and $x_3 > 0$ are
empty. There is a opening $D$ in the screen, that is $D$ is an open subset of
the plane $x_3 = 0$, not necessary connected (e.g., two slits). Finally, let
the screen be stationary, that is the geometry is time independent.

For the interaction between the particle and the screen, we assume: Inside the
half-spaces $x_3 < 0$, $x_3 > 0$, the wave function $\psi(\vec{x},t)$ of $S$
satisfies the free Schr\"{o}dinger equation,
\begin{equation}\label{schrodfree} i\hbar\frac{\partial
\psi(\vec{x},t)}{\partial t} = -\frac{\hbar^2}{2\mu}\left(\frac{\partial^2
\psi(\vec{x},t)}{\partial x_1^2} + \frac{\partial^2 \psi(\vec{x},t)}{\partial
x_2^2} + \frac{\partial^2 \psi(\vec{x},t)}{\partial x_3^2}\right)\ .
\end{equation}

Let us denote the part of the solution $\psi(\vec{x},t)$ in the left
half-space $x_3 < 0$ by $\psi_{\text{i}}(\vec{x},t)$ and in the right
half-space by $\psi_{\text{traf}}(\vec{x},t)$. Let
$\psi_{\text{i}}(\vec{x},t)$ be the $x_3 < 0$-part of a wave packet with $p_3
> 0$,
\begin{equation}\label{wavepack} \psi(\vec{x},t) =
\left(\frac{1}{2\pi\hbar}\right)^{3/2}\int_{{\mathbb
R}^3}d^3p\,\tilde{\psi}(\vec{p})\exp\left[\frac{i}{\hbar}\left(-\frac{|\vec{p}|^2}{2\mu}t
+ \vec{p}\cdot\vec{x}\right)\right]\ ,
\end{equation} where $\tilde{\psi}(\vec{p})$ is a rapidly decreasing function
(see \cite{RS}, p.\ 133) with $\tilde{\psi}(\vec{p}) = 0$ for all $p_3 \leq
0$, and let, for any fixed (finite) time, function
$\psi_{\text{traf}}(\vec{x},t)$ is rapidly decreasing.

At the points of the screen, the wave function is discontinuous. From the
left, the boundary values
$$
\lim_{x_3\rightarrow -0} \psi(\vec{x},t) = \lim_{x_3\rightarrow 0}
\psi_{\text{i}}(\vec{x},t)\ ,\quad \lim_{x_3\rightarrow -0} \frac{\partial
\psi}{\partial x_3}(\vec{x},t) = \lim_{x_3\rightarrow 0} \frac{\partial
\psi_{\text{i}}}{\partial x_3}(\vec{x},t)\ ,
$$
are determined by the solution $\psi_{\text{i}}(\vec{x},t)$. From the right,
\begin{equation}\label{bcscreen0} \lim_{x_3\rightarrow 0}
\psi_{\text{traf}}(\vec{x},t) = 0\ ,\quad \lim_{x_3\rightarrow 0}
\frac{\partial \psi_{\text{traf}}}{\partial x_3}(\vec{x},t) = 0
\end{equation} for $(x_1,x_2) \not\in D$ and
\begin{equation}\label{bcscreen1} \lim_{x_3\rightarrow 0}
\psi_{\text{traf}}(\vec{x},t) = \lim_{x_3\rightarrow -0} \psi(\vec{x},t)\
,\quad \lim_{x_3\rightarrow 0} \frac{\partial \psi_{\text{traf}}}{\partial
x_3}(\vec{x},t) = \lim_{x_3\rightarrow -0} \frac{\partial \psi}{\partial
x_3}(\vec{x},t)
\end{equation} for $(x_1,x_2) \in D$.

This expresses the notion that all particles arrive at the screen from the
left and those that hit the screen are absorbed by the screen and cannot
reappear.

The mathematical problem defined by the above assumptions can be solved by the
same method as the diffraction problem in optics can (see \cite{BW}, Section
8.3.1)\footnote{The author is indebted to Pavel Kurasov for clarifying this
point.} even if the wave equation is a rather different kind of differential
equation than the Schr\"{o}dinger equation. Indeed, for a monochromatic wave,
$$
\psi(\vec{x},t) = \exp\left(-\frac{i}{\hbar}Et \right)\Psi(\vec{x})\ ,
$$
Eq.\ (\ref{schrodfree}) implies
$$
\bigtriangleup \Psi(\vec{x}) + k^2 \Psi(\vec{x}) = 0\ ,
$$
where
$$
k^2 = \frac{2\mu E}{\hbar^2}\ ,
$$
which coincides with Helmholtz equation (\cite{BW}, p.\ 375). The solution of
Helmholtz equation in the half-space $x_3 > 0$ given by Fresnel-Kirchhof
diffraction formula (\cite{BW}, p.\ 380) then leads to the general solution
$\psi_{\text{traf}}(\vec{x},t)$ (which is a Fourier integral of monochromatic
waves defined by $\tilde{\psi}(\vec{p})$ of Eq. (\ref{wavepack})) that
satisfies the required boundary conditions. Hence, the solution exists and is
unique.

We can define absorption, ${\mathrm P}_{\text{abs}}$, and transmission,
${\mathrm P}_{\text{tra}}$, probabilities for the screen as follows:
\begin{equation}\label{tranfprob} {\mathrm P}_{\text{tra}} =
\lim_{t\rightarrow\infty}\int_{{\mathbb R}^2}d^2x\,\int_0^\infty dx_3
|\psi_{\text{traf}}(\vec{x},t)|^2
\end{equation} and
$$
{\mathrm P}_{\text{abs}} = 1 - {\mathrm P}_{\text{tra}}\ .
$$
This is based on the idea that the initial rapidly decreasing wave packet will
leave the left half-space completely for $t\rightarrow\infty$.

Function $\psi_{\text{traf}}(\vec{x},t)$ is not normalised and its norm is
${\mathrm P}^2_{\text{tra}} < 1$. Hence, the model defines a dynamics that is
not unitary. This is clearly due to the incompleteness of the model: particles
that hit the screen are absorbed and this part of the process was ignored
above. Let us give a short account of the physics of absorption. Let screen
$B_q$ be a macroscopic quantum system with Hilbert space ${\mathbf H}^B$ (a
real screen is somewhat thicker than a plane, but we just construct a
model). The process of disappearance of a quantum system $S$ in a macroscopic
body $B_q$ can be decomposed into three steps. First, $S$ is prepared in a
state that has a separation status so that a further preparation or
registration (in which the screen participates) can be made. Second, such $S$
enters $B_q$ and ditch most of its kinetic energy somewhere inside
$B_q$. Third, the energy passed to $B_q$ is dissipated and distributed
homogeneously through $B_q$ in a process aiming at thermodynamic
equilibrium. Then, system $S$ ceases to be an object and it does not possess
any individual state of its own after being absorbed if there are any particle
of the same type within $B_q$, as it has been explained in
\cite{incomplete}. It loses its separation status. Even if, originally, no
particle of the same type as $S$ is within $B_q$, in the course of the
experiment, $B_q$ will be polluted by many of them. The body is assumed to be
a perfect absorber so that $S$ does not leave it. Thus, the screen is assumed
to be {\em ideal}: every particle that arrives at it is either absorbed or
goes through the opening.

It is important that the absorption process is (or can be in principle)
observable. For instance, the increase of the temperature of $B_q$ due to the
energy of the absorbed particles can be measured. That is, either a single
particle $S$ has enough kinetic energy to cause an observable temperature
change, or there is a cumulative effect of more absorbed particles. More
precisely, suppose that the energy $E^{\mathcal S}$ of the absorbed particle
is small,
\begin{equation}\label{limclscreen} E^{\mathcal S} < \Delta E^{\mathcal B}\ ,
\end{equation} where $\Delta E^{\mathcal B}$ is the variance of the screen
energy in the initial state of the screen so that it would seem that the
absorption could not change the classical state of the screen. However, after
a sufficient number of absorptions, the total change of the energy will
surpass the limit (\ref{limclscreen}) so that the average change of the screen
energy due to one absorption is well defined. In any case, the initial and
final states of $B_q$ cannot be described by wave functions and they differ by
their classical properties from each other, e.g.\ by the temperature (see also
\cite{hajicekC3}).

Let us now try to complete the model including the process of absorption by
writing the initial state as a linear combination of the absorbed and the
transmitted ones. We define a function $\psi_{\text{trai}}(\vec{x},t)$ for
$x_3 < 0$ as the solution of Schr\"{o}dinger equation (\ref{schrodfree})
satisfying the boundary conditions
\begin{equation}\label{bcscreen0-} \lim_{x_3\rightarrow 0}
\psi_{\text{trai}}(\vec{x},t) = 0\ ,\quad \lim_{x_3\rightarrow 0}
\frac{\partial \psi_{\text{trai}}}{\partial x_3}(\vec{x},t) = 0
\end{equation} for $(x_1,x_2) \not\in D$ and
\begin{equation}\label{bcscreen1-} \lim_{x_3\rightarrow 0}
\psi_{\text{trai}}(\vec{x},t) = \lim_{x_3\rightarrow 0}
\psi_{\text{i}}(\vec{x},t)\ ,\quad \lim_{x_3\rightarrow 0} \frac{\partial
\psi_{\text{trai}}}{\partial x_3}(\vec{x},t) = \lim_{x_3\rightarrow 0}
\frac{\partial \psi_{\text{i}}}{\partial x_3}(\vec{x},t)
\end{equation} for $(x_1,x_2) \in D$.

Then, the pair of functions $\psi_{\text{trai}}(\vec{x},t)$ and
$\psi_{\text{traf}}(\vec{x},t)$ define a $C^1$ solution to the Schr\"{o}dinger
equation in the whole space as if the screen did not exist. Let us denote this
function by $\sqrt{{\mathrm
P}_{\text{tra}}}\psi_{\text{tra}}(\vec{x},t)$. Then,
$\psi_{\text{tra}}(\vec{x},t)$ is a normalised solution running from the left
to the right and vanishing in the left-hand half-space for large times.

Finally, let us define function $\psi_{\text{abs}}(\vec{x},t)$ in the
left-hand half-space by
\begin{equation}\label{decomp1} \psi(\vec{x}) = c_{\text{tra}}
\psi_{\text{tra}}(\vec{x}) + c_{\text{abs}} \psi_{\text{abs}}(\vec{x})\ ,
\end{equation} where $c_{\text{tra}} = \sqrt{{\mathrm P}_{\text{tra}}}$,
$c_{\text{abs}} = \sqrt{1 - {\mathrm P}_{\text{tra}}}$ and
$\psi_{\text{tra}}(\vec{x},t)$ is a normalised wave function of the part that
will be left through and $\psi_{\text{abs}}(\vec{x})$ that that will be
absorbed by $B_q$. Indeed, the two wave functions $\psi_{\text{tra}}$ and
$\psi_{\text{abs}}$ must be orthogonal to each other because their large-time
evolution gives $\psi_{\text{abs}} = 0$ in the right-hand half space and
$\psi_{\text{tra}} = 0$ in the left-hand half space.

Decomposition (\ref{decomp1}) is determined by the nature of $B_q$: for a
polariser, these are the two orthogonal polarisation states, and for a simple
screen consisting of a wall with an opening, these can be calculated from the
geometry of $B_q$ and the incoming beam.

The initial state of $B_q$ is a high-entropy one (see \cite{hajicekC3}). It
is, therefore, described by a state operator ${\mathsf T}_{\text{i}}$. Then
the initial state for the evolution of the composite is
$$
\bar{\mathsf T}_{\text{i}} = N^2_{\text{exch}} \bar{\mathsf
\Pi}_S(|\psi_{\text{i}}\rangle \langle\psi_{\text{i}}| \otimes {\mathsf
T})\bar{\mathsf \Pi}_S\ ,
$$
where $N^2_{\text{exch}} = tr\Bigl(\bar{\mathsf \Pi}_S(|\psi\rangle
\langle\psi| \otimes {\mathsf T})\bar{\mathsf \Pi}_S\Bigr)$ and $\bar{\mathsf
\Pi}_S$ is the (anti-)symmetrization over all particles indistinguishable from
$S$ (see \cite{incomplete}) within the composite system $S + B_q$ (we leave
open the question of whether they are fermions or bosons---thus we make a more
general theory than that of the previous subsection). It is an operator on the
Hilbert space $\bar{\mathsf \Pi}_S({\mathbf H} \otimes {\mathbf
H}^B)$. Further steps are analogous to those for the absorption of the
registered system in the photo-emulsion ${\mathcal D}$ that has been analysed
in more details in the previous section and we can skip the details here.

Let the evolution of the composite $S + B_q$ be described by operator
$\bar{\mathsf U}$. It contains the absorption and dissipation process in
$B_q$. $\bar{\mathsf U}$ is a unitary operator on the Hilbert space ${\mathbf
H} \otimes {\mathbf H}^B$ that commutes with projection $\bar{\mathsf \Pi}_S$
(see Section 5 of \cite{incomplete}) so that it leaves $\bar{\mathsf
\Pi}_S({\mathbf H} \otimes {\mathbf H}^B)$ invariant and defines an operator
in Hilbert space $\bar{\mathsf \Pi}_S({\mathbf H} \otimes {\mathbf H}^B)$ of
the composite. It is independent of the choice of the initial state. After the
process is finished, we obtain
$$
\bar{\mathsf T}_{\text{f}} = N^2_{\text{exch}}\bar{\mathsf \Pi}_S\bar{\mathsf
U} (|\psi_{\text{i}}\rangle \langle\psi_{\text{i}}| \otimes {\mathsf
T}_{\text{i}})\bar{\mathsf U}^\dagger\bar{\mathsf \Pi}_S\ .
$$

Using decomposition (\ref{decomp1}), we can write
\begin{multline}\label{formscreenf} \bar{\mathsf T}_{\text{f}} =
N^2_{\text{exch}}c_{\text{abs}}c^*_{\text{abs}}\bar{\mathsf \Pi}_S\bar{\mathsf
U} (|\psi_{\text{absi}}\rangle \langle\psi_{\text{absi}}| \otimes {\mathsf
T}_{\text{i}})\bar{\mathsf U}^\dagger\bar{\mathsf \Pi}_S \\ +
N^2_{\text{exch}}c_{\text{tra}}c^*_{\text{tra}}\bar{\mathsf \Pi}_S\bar{\mathsf
U} (|\psi_{\text{trai}}\rangle \langle\psi_{\text{trai}}| \otimes {\mathsf
T}_{\text{i}})\bar{\mathsf U}^\dagger\bar{\mathsf \Pi}_S \\ +
N^2_{\text{exch}} c_{\text{tra}}c^*_{\text{abs}}\bar{\mathsf
\Pi}_S\bar{\mathsf U} (|\psi_{\text{trai}}\rangle \langle\psi_{\text{absi}}|
\otimes {\mathsf T}_{\text{i}})\bar{\mathsf U}^\dagger\bar{\mathsf \Pi}_S \\ +
N^2_{\text{exch}}c_{\text{abs}}c^*_{\text{tra}}\bar{\mathsf \Pi}_S\bar{\mathsf
U} (|\psi_{\text{absi}}\rangle \langle\psi_{\text{trai}}| \otimes {\mathsf
T}_{\text{i}})\bar{\mathsf U}^\dagger\bar{\mathsf \Pi}_S\ .
\end{multline} The first term describes the process that starts with state
$\psi_{\text{absi}}$. Thus, $S$ does not reappear at the end and the result is
an excited state $\bar{\mathsf T}'_{\text{f}}$ of the screen that has absorbed
$S$. The second term represents the evolution that starts with $S$ in the
state $\psi_{\text{trai}}$. Then the screen remains in its initial state
${\mathsf T}_{\text{i}}$ and $S$ reappears in state
$\psi_{\text{traf}}$. Hence,
$$
\bar{\mathsf T}_{\text{f}} = \bar{\mathsf T}_{\text{end}1} + \bar{\mathsf
T}_{\text{end}0}\ ,
$$
where
$$
\bar{\mathsf T}_{\text{end}1} = |c_{\text{abs}}|^2 \bar{\mathsf T}'_{\text{f}}
+ |c_{\text{tra}}|^2 |\psi_{\text{traf}}\rangle \langle \psi_{\text{traf}}|
\otimes {\mathsf T}_{\text{i}}\ .
$$
State $\bar{\mathsf T}_{\text{end}1}$ is a convex combination of two states
that differ from each other by their classical properties while
$$
tr(\bar{\mathsf T}_{\text{end}0}) = 0\ .
$$

We can now argue in analogy with the previous section: RCU interpretation
suggests together with the observation that only the first two terms describe
the state of the composite after each individual individual process and the
true end state is not just a convex combination but a proper mixture:
\begin{equation}\label{screenf} \bar{\mathsf T}_{\text{true}f} = {\mathrm
P}_{\text{tra}} |\psi_{\text{traf}}\rangle \langle\psi_{\text{traf}}| \otimes
{\mathsf T}_{\text{i}}\ +_s\ {\mathrm P}_{\text{abs}} \bar{\mathsf
T}'_{\text{f}}\ ,
\end{equation} where $+_s$ denotes a proper mixture, see \cite{hajicekC3}. The
transformation from $\bar{\mathsf T}_f$ to $\bar{\mathsf T}_{\text{true}f}$
such a mixture is our version of state reduction as in Section 2.1.

Again, the state reduction is not a unitary transformation: First, the
non-diagonal terms in (\ref{formscreenf}) have been erased. Second, we have
also assumed that state $\psi_{\text{traf}}$ is the state of $S$ that has been
{\em prepared} by the screening. This means for us that it is a real state
with a separation status. Hence, operator ${\mathsf \Pi}_S$ can be left out in
Formula (\ref{screenf}), see Section 5 of \cite{incomplete}. This is, of
course, another violation of unitarity.

The disappearance of $S$ in $B_q$, as well as the disappearance of $S$ in the
photo-emulsion ${\mathcal D}$ described in the previous section, is a physical
process that have a definite time and place. This suggests that the state
reduction occurs at the time and the place of the possible absorption of the
particle in $B_q$ or ${\mathcal D}$. The possible absorption had to be viewed
as a part of the whole process even in the case that an individual particle is
not absorbed but goes through. Indeed, that an individual particle goes
through is only a result of the state reduction, which is a change from the
linear superposition of the transition and the absorption states.

\section{The structure of meters} Here, we extend some ideas of Section 2.1 on
Stern-Gerlach apparatus to all meters with the aim to improve the
understanding of registrations. Most theoretical descriptions of meters that
can be found in the literature are strongly idealised (see, e.g.,
\cite{BLM,WM}): the meter is an arbitrary quantum system with a ``pointer''
observable. We are going to give a more elaborated picture and distinguish
between {\em fields, screens, ancillas} and {\em detectors} as basic
structural elements of meters.

Screens have been dealt with in Section 2.2. It is also more or less clear
what are fields: for example, in the Stern-Gerlach experiment, the beam is
split by an inhomogeneous magnetic field. In some optical experiments, various
crystals are used that make possible the split of different polarisations or
the split of a beam into two mutually entangled beams such as by the
down-conversion process in a crystal of KNbO$_3$ \cite{MW}. The corresponding
crystals can also be considered as fields. In any case, the crystals and
fields are macroscopic systems the (classical) state of which is not changed
by the interaction with the registered system.

In many modern experiments, in particular in non-demolition and weak
measurements, but not only in these, the following idea is employed. The
registered system $S$ interacts first with an auxiliary quantum system $A$
that is prepared in a suitable state. After $S$ and $A$ become entangled, $A$
is subject to further registration and, in this way, some information on $S$
is revealed. Subsequently, further measurements on $S$ can but need not be
made. The state of $S$ is influenced by the registration of $A$ just because
of its entanglement with $A$. Such auxiliary system $A$ is usually called {\em
ancilla} (see, e.g., \cite{peres}, p.\ 282).

Finally, important parts of meters are {\em detectors}. Indeed, even a
registration of an ancilla needs a detector. It seems that any registration on
microscopic systems has to use detectors in order to make features of
microscopic systems visible to humans. Detector is a large system that changes
its (classical) state during the interaction with the registered
system. ``Large'' need not be macroscopic but the involved number of particles
ought to be at least about $10^{10}$. For example, the photo-emulsion grain or
nanowire single photon detector (see, e.g., \cite{natarayan}) are large in
this sense. A criterion of being large is that the system has well-defined
thermodynamic states so that the thermodynamics is a good approximation for
some aspects of its behaviour.

For example, in the so-called cryogenic detectors \cite{stefan}, $S$
interacts, e.g., with superheated superconducting granules by scattering off a
nucleus in a granule. The resulting phonons induce the phase transition from
the superconducting to the normally conducting phase. The detector can contain
very many granules (typically $10^9$) in order to enhance the probability of
such scattering if the interaction between $S$ (a weakly interacting massive
particle, neutrino) and the nuclei is very weak. Then, there is a solenoid
around the vessel with the granules creating a strong magnetic field. The
phase transition of only one granule leads to a change in magnetic current
through the solenoid giving a perceptible electronic signal.

Modern detectors are constructed so that their signal is electronic. For
example, to a scintillation film, a photomultiplier is attached (as in
\cite{tono}). We assume that there is a signal collected immediately after the
detector changes its classical state, which we call {\em primary}
signal. Primary signal may still be amplified and filtered by other electronic
apparatuses, which can transform it into the final signal of the detector. For
example, the light signal of a scintillation film in the interference
experiment of \cite{tono} is a primary signal. It is then transformed into an
electronic signal by a photocathode and the resulting electronic signal is
further amplified by a photomultiplier.

A detector contains {\em active volume} ${\mathcal D}$ and {\em signal
collector} ${\mathcal C}$ in thermodynamic state of metastable
equilibrium. Notice that the active volume is a physical system, not just a
volume of space. For example, the photo-emulsion or the set of the
superconducting granules are active volumes. Interaction of the detected
systems with $\mathcal D$ triggers a relaxation process leading to a change of
the classical state of the detector---the {\em detector signal}. For some
theory of detectors, see, e.g., \cite{leo,stefan}.

What is the difference between ours and the standard ideas on detectors? The
standard ideas are, e.g., stated in (Ref.\ \cite{peres} p.\ 17) with the help
of the Stern-Gerlach example:
\begin{quote} The microscopic object under investigation is the magnetic
moment $\mathbf \mu$ of an atom.... The macroscopic degree of freedom to which
it is coupled in this model is the centre of mass position $\mathbf r$... I
call this degree of freedom {\em macroscopic} because different final values
of $\mathbf r$ can be directly distinguished by macroscopic means, such as the
detector... From here on, the situation is simple and unambiguous, because we
have entered the macroscopic world: The type of detectors and the detail of
their functioning are deemed irrelevant.
\end{quote} The root of such notion of detectors may be found among some ideas
of the grounding fathers of quantum mechanics. For example, Ref.\
\cite{pauli}, p. 64, describes a measurement of energy eigenvalues with the
help of scattering similar to Stern-Gerlach experiment, and Pauli explicitly
states:
\begin{quote} We can consider the centre of mass as a 'special' measuring
apparatus\ldots
\end{quote}

In these statements, no clear distinction is made between ancillas and
detectors: indeed, the centre-of-mass position above can be considered as an
ancilla. However, such a distinction can be made and it ought to be made
because it improves our understanding of registrations. To be suitable for
this aim, we have slightly modified the current notions of detector and
ancilla. Our detectors are more specific than what is often assumed.

The foregoing analysis motivates the following hypothesis.
\begin{assump}\label{aspointerh} Any meter for microsystems must contain at
least one detector and every reading of the meter can be identified with a
primary signal from a detector. The state reduction required by realism and
observational evidence on measurements takes place in detectors and screens.
\end{assump} A similar hypothesis has been first formulated in
\cite{hajicek2}. Assumption \ref{aspointerh} makes the reading of meters less
mysterious.

\section{Two hypotheses on state reduction} Here, we study the form of state
reduction and the objective circumstances with which it is connected.
\begin{assump}\label{assh} Let ${\mathcal O}$ be an object (such as a
detector) with classical model $O_c$ and quantum model $O_q$. Let the standard
unitary evolution describing some process in which $O_q$ takes part results in
an end state of the form:
\begin{equation}\label{Tf} \bar{\mathsf T}_f = \sum_{k=1}^n {\mathrm P}_k
\bar{\mathsf T}_k + \bar{\mathsf T}_{\text{end}0}\ ,
\end{equation} where $\bar{\mathsf T}_k$ are states of $O_q$ such that each is
associated with a classical state of $O_c$ and these classical states are
different for different $k$'s. The coefficients satisfy ${\mathrm P}_k > 0$
for $k = 1,\ldots,n$ and $\sum_k {\mathrm P}_k = 1$. $\bar{\mathsf
T}_{\text{end}0}$ is a s.a.\ operator with trace 0. Then, the standard unitary
evolution must be corrected so that $\bar{\mathsf T}_f$ is replaced by
\begin{equation}\label{Tend} \bar{\mathsf T}_{\text{end}} = \sum_{k=}^n +_s
{\mathrm P}_k \bar{\mathsf T}_k\ ,
\end{equation} the proper mixture of states $\bar{\mathsf T}_k$.
\end{assump} Assumption \ref{assh} is applicable to those unitary evolutions
that have an end state of the form (\ref{Tf}). However, classical objects may
have some properties that make such a form to be a general case. For example,
it may be impossible for a quantum model of a classical object to be in a
convex combination of states, one of which is associated with a classical
state and the other not having classical properties or in a state equal to two
different convex compositions so that the two sets of classical states defined
by the two compositions are different from each other. This seems to follow
from the classical realism described in \cite{hajicekC3}.

To illustrate the difference to an ordinary convex decomposition, let us
consider an arbitrary normalised state vector $\Phi$ of some quantum
system. Such a state can be decomposed into two orthonormal vectors in an
infinite number of different ways, for example,
$$
\Phi = c_1\Phi_1 + c_2\Phi_2 = d_1\Psi_1 + d_2\Psi_2\ .
$$
Then
$$
|\Phi\rangle \langle \Phi| = |c_1|^2 |\Phi_1\rangle \langle \Phi_1| + |c_2|^2
|\Phi_2\rangle \langle \Phi_2| + c_1c^*_2|\Phi_1\rangle\langle \Phi_2| +
c_2c^*_1|\Phi_2\rangle\langle \Phi_1|
$$
and
$$
|\Phi\rangle \langle \Phi| = |d_1|^2 |\Psi_1\rangle \langle \Psi_1| + |d_2|^2
|\Psi_2\rangle \langle \Psi_2| + d_1d^*_2|\Psi_1\rangle\langle \Psi_2| +
d_2d^*_1|\Psi_2\rangle\langle \Psi_1|
$$
are two different decompositions of state $|\Phi\rangle \langle \Phi|$ that
have the form of (\ref{Tf}).

Assumption \ref{assh} defines a rule that determines the correction to unitary
evolution {\em uniquely} in a large class of scattering and registration
processes (see \cite{hajicek4,hajicek5}). We leave the detailed questions of
applicability of Assumption \ref{assh} open to future investigations in the
hope that the approach that it suggests is more or less clear.

Both detectors and screens, where the state reductions occur, are mezzo- or
macroscopic (for example, the emulsion grains can be considered as
mezzoscopic), but there are processes of interaction between microscopic and
macroscopic objects, the standard quantum description of which gives always a
unique classical end state of the macroscopic part. For example, the
scattering of neutrons by ferromagnetic crystals in which the crystal remains
in the same classical state during the process of scattering. In such
processes, Assumption \ref{assh} implies no state reductions. It is the
structure of the final quantum state that makes the difference: for a state
reduction, the standard quantum evolution had to give a convex combination of
states that differ in their classical properties.

What is the cause of the change $\bar{\mathsf T}_f$ into $\bar{\mathsf
T}_{\text{end}}$? For example, the detector that detects microsystem $S$
achieves the signal state so that $S$ interacts with its active volume
${\mathcal D}$ and the state of $S + {\mathcal D}$ dissipates, which leads to
a loss of separation status of $S$. A similar process runs in a screen that
absorbs $S$. The dissipation is necessary to accomplish the loss. The
dissipation process does not have anything mysterious about it. It can be a
usual thermodynamic relaxation process in a macroscopic system or a similar
process of the statistical thermodynamics generalised to nano-systems (see,
e.g., \cite{horodecki}). $S$ might be the registered object or an ancilla of
the original experiment. In all such cases, state $\bar{\mathsf
T}_{\text{end}}$ originates in a process of relaxation triggered by $S$ in
${\mathcal O}$ and accompanied by the loss of separation status of $S$. This
motivates the following hypothesis:
\begin{assump}\label{asavh} The cause of the state reduction postulated by
Assumption \ref{assh} is an uncontrollable disturbance due to a loss of
separation status.
\end{assump} The loss of separation status is an objective process and the
significance of Assumption \ref{asavh} is that it formulates an objective
condition for the applicability of an alternative kind of dynamics.

Actually, the assumption that a measuring process disturbs the measured system
in an uncontrollable way and that this is the cause of the state reduction is
not new (see, e.g., \cite{messiah}, Section 4.3.1). What we add to it is just
the role of separation-status loss.

The three hypotheses \ref{aspointerh}, \ref{assh} and \ref{asavh} form a basis
of our theory of state reduction. They generalise some empirical experience,
are rather specific and, therefore, testable. That is, they cannot be
disproved by purely logical argument but rather by an experimental
counterexample. For the same reason, they also show a specific direction in
which experiments ought to be proposed and analysed: if there is a state
reduction, does then a loss of separation status take part in the process?
What system loses its status? How the loss of the status can lead to state
reduction?

In fact, our theory remains rather vague with respect to the last question in
that it suggests no detailed model of the way from a separation status change
to a state reduction. Such a model would require some new physics and we
believe that hints of what this new physics could be will come from attempts
to answer the above questions by suitable experiments.

Many examples and models of state reduction were studied in papers
\cite{hajicek4,hajicek5} that are based on the old definition of separation
status. A reformulation of the examples for the new definition given here is
more or less straightforward.

\section{Conclusion} The basic idea on the structure of meters and the role of
detectors as explained in \cite{hajicek2,hajicek4} has been adapted to the new
definition \cite{incomplete} of separation status. Three main improvements
resulted.

First, the restriction of state reduction to registration processes has been
removed and a general theory of state reduction has been introduced and
explained by the example of screening. For such generalisation, a clear
distinction between scattering and partial and complete absorption of a
particle is necessary and it is provided by the presence or absence of
dissipation.

Second, the restriction of \cite{hajicek2,hajicek4} to macroscopic meters can
be abandoned because dissipation processes are possible also in much smaller
detectors. Thus, our theory becomes applicable to many modern experiments.

Third, papers \cite{hajicek2,hajicek4} used the notion of separation status in
an incorrect way because the their misleading limitation to the geometrical
aspects of the experimental arrangements. The generalised notion of separation
status enabled a formulation of the theory of state reduction in a way that is
independent of representation it so that it is covariant with respect to
unitary transformations.

Finally, an example based on superconducting rings \cite{LS} seems to suggests
that an experimental check of the theory is possible. In summary, a better
understanding of the notion of state reduction has resulted.

\section*{Acknowledgements} The author is indebted to Nicolas Gisin, Stefan
J\'{a}no\v{s}, Petr Jizba and Ji\v{r}\'{\i} Tolar for useful discussions.

\end{document}